\documentstyle[twocolumn,prl,aps]{revtex}
\begin{document}
\draft
\title{Rigorous results on a first order phase transition in
antiferromagnetic spin-1/2 coupled chains}
\author{Y. Xian \cite{email}}
\address{Department of Mathematics,\\
University of Manchester Institute of Science and Technology (UMIST),\\
P.O. Box 88, Manchester M60 1QD, UK\\
{\rm (November 29, 1994)}}
\maketitle

\begin{abstract}
Some rigorous results are presented for a first-order quantum phase
transition between the dimerized state and Haldane-type state (i.e., a
state similar to the ground state of the one-dimensional spin-1
Heisenberg chain) in the spin-1/2 coupled chains with
nearest-neighbour and next-nearest-neighbour Heisenberg interactions.
Also presented are the exact excited states in both phases. A partial
phase diagram of the general spin-1/2 coupled chains is discussed.
\end{abstract}


Spin-1/2 coupled chains have attracted attention for the last few
years \cite {scala,rice,noak} partly because of their relevance to
such materials as (VO)${}_2$P${}_2$O${}_7$ \cite{exp1} and
Sr${}_2$Cu${}_4$O${}_6$ \cite{exp2}, and partly because of their
intrinsic theoretical interest. For the isotropic coupled chains with
only nearest-neighbour couplings, current consensus is that the system
has a nonzero energy gap separating the ground and low-lying excited
states, and the spin-spin correlation length of the ground state is
finite \cite{scala,rice,noak}. No long-range order has been reported.

It is well known that, for one-dimensional spin chain systems, the
spontaneous dimerized phase (the perfect dimer state being defined as
every two adjacent atoms forming a spin-singlet pair in the spin-1/2
or spin-1 chains) \cite{dimer} and Haldane-type phase (spin-1 chain)
\cite{spin1} are both possible. It is also well known that these two
phases both exhibit nonzero gap and finite correlation length. The
main difference between the two phases lies in the fact that the
Haldane-type state is nondegenerate and translationally invariant with
a hidden string order \cite{aklt,white}, while the spontaneous
dimerized state of a single chain is doubly degenerate with
translational symmetry broken. For the system of spin-1/2 coupled
chains, it is therefore natural to consider the possible existence of
these two phases in despite of the dimerized phase being
non-spontaneous and to study corresponding phase transitions if the
two phases do exist. In fact, spin-1/2 coupled chains as a subject of
research have a longer history \cite{nijss,solyom}. The main interest
there was to construct effective one-dimensional chains with larger
spin moments from spin-1/2 coupled chains. Many interesting results
were produced \cite{nijss,solyom}. In particular, it was found that
spin-1/2 coupled chains can have the same ground state as that of the
one-dimensional spin-1 Heisenberg chain.

Because of their relevance to the experiments, the general spin-1/2
coupled chains are important models on their own right. In this
article, we study spin-1/2 coupled chains which clearly exhibit both
the dimerized phase and Haldane-type phase when the coupling
parameters vary. Specifically, we consider spin-1/2 coupled chains
with nearest-neighbour and next-nearest-neighbour Heisenberg
interactions. The spin system consists of $N$ pairs of $S=1/2$ atoms,
with interactions described by the following Hamiltonian,
\begin{eqnarray}
  H&=&\sum_{r=1}^N [J{\bf S}_1^r\cdot{\bf S}_2^r +
    J'({\bf S}_1^r\cdot{\bf S}_1^{r+1}
        + {\bf S}_2^r\cdot{\bf S}_2^{r+1}) \nonumber \\
   &+& J''({\bf S}_1^r\cdot{\bf S}_2^{r+1} +
    {\bf S}_2^r\cdot{\bf S}_1^{r+1})],
\end{eqnarray}
where the rung index $r$ denotes each pair of spins with the spins on
the top chain denoted as ${\bf S}_1(r)$ and the spins on the bottom
chain denoted as ${\bf S}_2(r)$; $J$ is the coupling constant across
each rung, $J'$ is the nearest-neighbour coupling constant along each
chain, and $J''$ is the next-nearest-neighbour (diagonal) coupling
constant. We mainly discuss the antiferromagnetic region ($J,J',J''
\ge 0$). Periodic boundary conditions are taken in this article,
namely, ${\bf S}_\mu^{N+1} = {\bf S}_\mu^1$, with $\mu = 1,2$. The
Hamiltonian of Eq. (1) at $J=0$ is the so-called composite spin model
\cite{solyom}.

We first consider a single rung with a pair of spin-1/2 atoms only.
The rung has four states: the singlet state with zero total spin,
namely $|0\rangle ={\frac 1{\sqrt{2}}}(|\uparrow \downarrow
\rangle -|\downarrow \uparrow \rangle )$ in the usual spin-1/2
notation; and the triplet states with total spin equal to 1, namely
$|1\rangle =|\uparrow \uparrow \rangle $, $|2\rangle =
{\frac 1{\sqrt{2}}}(|\uparrow \downarrow \rangle +|\downarrow
\uparrow \rangle )$, and $|3\rangle =|\downarrow
\downarrow \rangle $ respectively. In a matrix representation
\cite{parkinson,xian}, these four states are denoted by four column
matrices respectively, and the single spin operators ($S_\mu ^z$ and
$S_\mu ^{\pm }$ with $\mu =1,2$) can then be written as $4\times 4$
matrices. We then introduce the so-called composite operator $A_{nm}$
($n,m=0,1,2,3$), which is a $4\times 4$ matrix with a single nonzero
entry, namely $\langle n'|A_{nm}|m'\rangle =\delta _{nn'}\delta
_{mm'}$. The single spin operators can then be written as linear
summations of composite operators $A_{nm}$.  The meaning and
bosonization of these composite operators have been discussed in
detail in \cite{xian}. Briefly, $A_{00}\ (=1/4-{\bf S}_1\cdot {\bf
S}_2)$ is the singlet projection operator, $A_{0n}$ and $A_{n0}$
($n=1,2,3$) make transitions between the singlet and triplet states,
and $A_{nm} $ ($n,m=1,2,3$) make transitions among the triplet states.
$A_{nm}$ obey pseudo-spin algebra, $[A_{nm},A_{kl}] =
A_{nl}\delta_{mk} - A_{km}\delta_{ln}$.

In terms of these composite operators, $A_{nm}^r$, with the index $r$
denoting each rung in the coupled chains, the Hamiltonian of Eq. (1)
can be rewritten as
\begin{eqnarray}
  H &=&\sum_{r=1}^N\Bigl[J\left({1\over4}-A_{00}^r\right)
    + {1\over2}(J'-J'')H_{r,r+1} \nonumber \\
    &+& {\frac{1}{2}}(J'+J'') H'_{r,r+1}\Bigr],
\end{eqnarray}
where $H_{r,r+1} \equiv (A_{02}^r+A_{20}^r)
(A_{20}^{r+1}+A_{02}^{r+1}) + (A_{01}^r-A_{30}^r)
(A_{10}^{r+1}-A_{03}^{r+1}) + (A_{10}^r-A_{03}^r)
(A_{01}^{r+1}-A_{30}^{r+1})$ and $H'_{r,r+1} \equiv
(A_{11}^r-A_{33}^r) (A_{11}^{r+1}-A_{33}^{r+1}) +
(A_{21}^r+A_{32}^r)(A_{12}^{r+1}+A_{23}^{r+1}) +
(A_{12}^r+A_{23}^r)(A_{21}^{r+1}+A_{32}^{r+1})$. It is interesting to
compare this Hamiltonian with that of the one-dimensional spin-1/2
frustrated Heisenberg models (i.e., Eqs. (3.4)-(3.5) of Ref.
\cite{xian}). As we can see, by using the composite operators, the
algebra of the spin-1/2 coupled chains becomes quite transparent. It
is trivial to note that when $J'=0$ and $J''=0$, the ground state of
the system is simply the perfect dimer state $|D\rangle$ which is
given by the product of singlet states of all $N$ rungs, namely $|D
\rangle = \prod_{r=1}^N |0 \rangle_r$. More significantly, we notice
that the Hamiltonian of Eq. (2) consists of three parts, the first
part consisting only of the operators $A_{00}^r$ which is nonzero only
when acting on the singlet state of the $r$-th rung, the second part
consisting only of the operators $H_{r,r+1}$ which makes transition
between the singlet and triplet states, and finally the third part
consisting only of the operators $H'_{r, r+1}$ which is nonzero only
when acting on a state in which both the $r$-th and $(r+1)$-th rungs
are in the triplet states.

In fact, it is easy to prove that $H'_{r,r+1}$ is similar to the usual
spin-1 Heisenberg interaction \cite{solyom}, namely
\begin{equation}
 H'_{r,r+1}={\bf P}_r\cdot {\bf P}_{r+1},\ \ \ \ |{\bf P}|=1,
\end{equation}
where we define operator ${\bf P}_r$ by
\begin{eqnarray}
   P_r^z &=& A_{11}^r-A_{33}^r,\ \ \ \
      P_r^{+} = \sqrt{2}(A_{12}^r+A_{23}^r), \nonumber \\
   P_r^- &=& \sqrt{2}(A_{21}^r+A_{32}^r),
\end{eqnarray}
with the usual $SU(2)$ algebra, $[P^+,P^-]=2P^z$ and
$[P^z,P^\pm]=\pm P^\pm$.

At $J' = J''$, the second part of $H$ disappears. The Hamiltonian is
then reduced to
\begin{equation}
H_0 = \sum_{r=1}^N (-A_{00}^r + j'{\bf P}_r \cdot
        {\bf P}_{r+1}), \ \ \ \ J'' = J',
\end{equation}
where, for convenience, we have set $J'/J=j'$ and $J=1$, and where we
have ignored the constant term, $N/4$. We notice that operators
$A_{00}^r $ and ${\bf P}_r\cdot{\bf P}_{r+1}$ commute with one
another. We have therefore arrived at a Hamiltonian consisting of two
decoupled parts.  The first part, $-\sum_r A_{00}^r$, is trivial, with
no interaction between different rungs and with a gap value of 1
between the singlet and triplet states. The second part, $\sum_r {\bf
P}_r \cdot {\bf P}_{r+1}$, is similar to the spin-1 Heisenberg chain
with each rung in the coupled chains corresponding to each site in the
spin-1 chain (the only difference is appearance of a new type of
excitation to be discussed later). The spin-1 Heisenberg model has
been the focus of intensive investigations in the last decade and its
physical properties are now well understood. The state-of-the-art
calculations were recently performed by the density-matrix
renormalization-group technique \cite{white} and extremely accurate
results have been obtained. In particular, the long-range string
order, $g(\infty) = 0.374325096(2)$, is obtained by calculation of the
string correlation function $g(r)$ which is defined as
\begin{equation}
  g(r) = \left\langle P_0^z\left(
  \prod_{r'=1}^{r-1}{\rm e}^{i\pi P_{r'}^Z}\right)
  P_r^z\right\rangle.
\end{equation}

Since the two parts of $H_0$ are decoupled and since all excited
states of both parts have nonzero energy gaps (to be discussed later),
its ground state is either that of the first part, or that of the
second part, depending on the value of $j'$. The exact ground-state
energy of $H_0$ is therefore given by
\begin{equation}
  {\frac{E_g}{N}} = \left\{
  \begin{array}{l}
   -1\;\;\mbox{for}\;j'<j_c\;, \\
   -{j'\over j_c}\;\;\mbox{for}\;j'>j_c\;,
  \end{array}\right.
\end{equation}
where $j_c$ is the critical coupling constant, defined by $j_c \equiv
1/e_0=0.713529353310(1)$ with $e_0$ the ground-state energy per site
of the spin-1 Heisenberg chain \cite{white}. Clearly, the phase
transition at $j'=j_c$ is a first-order one because the first-order
energy derivative with respect to $j'$ is discontinuous at the critical
point.

We next discuss excitation states in two phases. For the dimerized
phase ($j'<j_c$), the excitations states can be easily obtained. For
convenience, we define $n$ singlet-to-triplet spin-flip (STSF) state
as a configuration in which $n$ rungs are in the triplet states, while
the remaining $(N-n)$ rungs are in the singlet state as in the dimer
ground state. If each flip is separated from each another in a
$n$-STSF configuration, the energy gap (i.e., total excitation energy
minus total ground-state energy) of the $n$-STSF state is simply
$\Delta E_d(n)=n$ with $3^n$-fold degeneracy. However, if $n$ flips
form a contiguous cluster, the excitation gap can be much lower, with
gap values given by $\Delta E_d(n)=n+E(n)j'$, where $E(n)$ are the
eigenvalues of the open-ended $n$ spin-1 Heisenberg chain. As well
known, the ground state of the open-ended spin-1 chain is singlet when
$n$ is even and triplet when $n$ is odd; the energy difference between
the singlet (triplet) ground state and next triplet (singlet) excited
state for a given $n$ decreases exponentially to zero in the large-$n$
limit and the exact ground-state energy was obtained for up to $n=14$
\cite{kennedy}. For the one-flip state, $E(1)=0$, therefore, $\Delta
E_d(1)=1$ with triplet degeneracy. For the contiguous 2-STSF cluster,
$\Delta E_d(2)=2(1-j'),2-j'$, and $2+j'$ for the singlet, triplet, and
quintuplet states respectively, with a minimum gap of about $0.573$
when $j' \rightarrow j_c$. (For 3- and 4-STSF clusters, the minimum
gaps as $j' \rightarrow j_c$ are $0.895$ and $0.685$ respectively). If
we consider a single STSF state as a (soliton) particle with spin
momentum equal to one, it is obvious these particles attract to one
another at low values of their total spin quantum numbers. For large
$n$ ($n>14$), $E(n) \equiv [-e_0(n-1)+ e']$, where $e'$ represents the
residual boundary effect (its magnitude is expected to be much smaller
than $e_0$ \cite{kennedy}), the excitation gap of the $n$ STSF cluster
state is then given by
\begin{equation}
  \Delta E_d(n) = n + [-e_0(n-1)+e']j', \ \ \ \ n\ \mbox{large}
\end{equation}
When $j' \rightarrow j_c$, $\Delta E_d(n) \rightarrow 1 + e'j_c$, a
number independent of $n$. Since $|e'| \ll 1$, $\Delta E_d(n) \approx
1$ for large $n$. When $n$ approaches $N$, it is easy to see that the
dimer ground state is unstable against the large $n$-STSF cluster
state when $j'>j_c$, as expected.

The excitations in the Haldane-type phase have been discussed for the
Hamiltonian of Eq. (1) at $J=0$ in Ref. \cite{solyom}. Here, we take
advantage of the recent accurate results for the spin-1 Heisenberg
chain \cite{white}, and focus on the excitations of Eq. (5) for
$j'>j_c$. As we know, the low-lying excited state of the spin-1 chain
is given by the one-magnon state with momentum $q=\pi$, and with the
well-known Haldane gap $\Delta = 0.41050(2)$ \cite{white}.
Multi-magnon states have larger gaps \cite{white,affleck}. We consider
here a new type of excitations where the spin-1 chain is embedded with
singlet rungs \cite{solyom}. Again, for convenience, we define $n$
triplet-to-singlet spin-flip (TSSF) configuration as a state in which
there are $n$ singlet rungs and the remaining $(N-n)$ rungs are in
their Haldane-ground-state configuration. We consider a single TSSF
configuration, which corresponds to the ground state of the $(N-1)$
spin-1 Heisenberg chain with two open ends enclosing the singlet rung.
Its total energy can be written as $E_h(1) = E(N-1)j'-1$, where, as
before, $E(N-1) = -e_0(N-2) + e'$. Hence, $\Delta E_h(1) =
(2e_0+e')j'- 1$. Since $|e'| \ll 1$, we obtain $\Delta E_h(1) \approx
1$ as $j' \rightarrow j_c$, a value larger than $0.4105j_c \approx
0.3$ of the one-magnon Haldane gap. Furthermore, from the numerical
calculations of the open-ended spin-1 chains \cite{white}, we note
that each of the two ends enclosing the singlet rung in the 1-TSSF
state has an effective $S=1/2$ spin with exponential decay of the
local spin moment away from the ends, and with decay length $\xi
\approx 6.03(1)$ (which is also the usual spin-spin correlation
length).  Therefore, the single TSSF configuration can be considered
as particle-like (soliton) state with a size of about 14 lattice
spacings (the core occupying two lattice spacings). We also notice
that the degeneracy of the single TSSF state is four, while the
one-magnon state of the spin-1 chain is triplet. (It is also not
difficult to obtain energy gaps of the 2-TSSF and other states and
these details will appear elsewhere). Similar to the $n$-STSF cluster
state in the dimer phase, the $n$-TSSF state has lowest energy when
$n$ flips are contiguous, with the gap
\begin{equation}
 \Delta E_h(n) = [(n+1)e_0+e']j'-n.
\end{equation}
We notice that when $j' \rightarrow j_c$, all clusters have the same
energy gap of $(1+e'j_c) \approx 1$, a number identical to that of the
(large) $n$-STSF cluster state of the dimer phase at the critical
point. When $n \rightarrow N$, Haldane-type phase is unstable against
formation of the large $n$-TSSF cluster state when $j'<j_c$, as
expected.

Finally, we discuss a partial phase diagram of the general Hamiltonian
of Eq. (1). At $J=J''=0$, the Hamiltonian reduces to the two decoupled
spin-1/2 Heisenberg chains which is integrable by Bethe {\it ansatz}
and which is critical with no gap. For $J''=0$, we apply the
(dimerized) valence-bond spin-wave theory developed by us\cite{xian}
via bosonization for the composite operators $A_{nm}^r$ in Eq. (2),
starting from the perfect dimer state $|D\rangle$ which is the exact
ground state at $J'=0$.  Good results for the ground and spin-wave
excited states are obtained for a small nonzero region of $J'/J$, but
the theory breaks down at $J'/J = 1/2$ and the dimerization order
(similar to that of the one-dimensional spin-1/2 frustrated chain
\cite{xian}) is nonzero in the region $0 < J'/J < 1/2$ \cite{xian2}.
This seems to suggest that the system of Eq.  (1) without
next-nearest-neighbour coupling (i.e., $J''=0$) is dimerized for a
nonzero region of $J'/J$ and there is a critical value of $J'/J$ at
which the dimerized system is unstable against transformation to other
phase(s). Combining with the rigorous results for $J''=J'$ presented
above, we propose a partial phase diagram as shown in Fig. 1. In Fig.
(a), $ABC$ is given by $J''=J'$ with $AB$ corresponds to the dimer
phase and $BC$ to the Haldane-type phase, and $BD$ is only for
guidance indicating the boundary between the dimerized phase and other
possible phase(s). We denote the dimerized region enclosed by $ABD$ as
I. $CE$ corresponding to the composite spin model is believed to show
Haldane-type phase except the critical point at $E$ \cite{solyom}. The
dashed line between $BD$ and $E$ is quite arbitrary at this moment. It
separates the Haldane-type phase region (denoted as II) and any other
possible phase region (denoted as III).

There are three possible scenarios for III: (a) The valence-bond
spin-wave theory is totally unreliable and the dimerized phase
persists from I to III; (b) The valence-bond spin-wave theory is
qualitatively correct and the whole region enclosed by $BCED$ (i.e.,
II plus III) is Haldane-type phase, including the isotropic model with
$J=J'$ and $J''=0$; (c) A new phase (such as spin liquid) appears in
III which also has a nonzero gap but without any long-ranged order
(dimerized or string order). We consider the first scenario most
unlikely because the (dimerized) valence-bond spin-wave theory is
expected to be at least qualitatively correct if the system has a
nonzero dimerization order. Results from analytical approximations and
numerical calculations \cite{scala,rice,noak} seem to suggest the
third scenario. However, since we have not seen any report on
calculations of the string order, we cannot rule out the second
scenario. Based on the rigorous results along $BC$ and numerical
results along $CE$, we tend to believe the second scenario. In any
case, it is interesting to apply powerful numerical techniques such as
the density-matrix renormalization-group \cite{noak} to calculate the
string order particularly in region III, using Eqs. (4) and (6).

To conclude this article, we should also point out that there are
other regions in Fig. 1 (a), where exact results can be obtained. For
example, we show in Fig. 1 (b) the complete phase diagram for the
Hamiltonian of Eq. (1) at $J''=J'$, where $GAB$ denotes the dimer
phase region, $BCF$ Haldane-type phase region, and $FG$ ferromagnetic
region. All of the three critical points, $G$, $B$ and $F$, which
separate the three phases, are first-order one.

\acknowledgments
I am grateful to R. F. Bishop and J. B. Parkinson for many useful
discussions.

\begin{figure}
\caption{A partial phase diagram for the spin-1/2 coupled chains
of Eq. (1). (a) Region I ($ABD$) is the dimerized phase, II ($BCE$) is
Haldane-type phase, III ($BED$) may be some new spin-liquid or
Haldane-type phase. $ABC$ is given by $J''=J'$ and the dashed line is
somewhat arbitrary. (b) The complete phase diagram when $J''=J'$, with
$GAB$ denoted the dimer phase, $BCF$ Haldane-type phase, and $FG$
ferromagnetic phase.}
\end{figure}

\end{document}